\documentclass[12pt]{iopart}

\usepackage{amssymb}
\usepackage{graphicx}
\begin{document}

\def\be{\begin{equation}}
\def\ee{\end{equation}}
\def\de{\Delta}
\def\sig{\Sigma}
\newcommand{\bea}{\begin{eqnarray}}
\newcommand{\eea}{\end{eqnarray}}
\newcommand{\eps}{\varepsilon}
\newcommand{\Vef}{V_{\mbox{\scriptsize eff}}}

\title {Energy density functional on a microscopic basis}
% {\bf Microscopic evaluation of the pairing gap}
\author{M Baldo}
\address{INFN, Sezione di Catania, 64 Via S.-Sofia, I-95125 Catania,
Italy}
\ead{baldo@ct.infn.it}
\author{L. Robledo}
\address{Dep. Fisica Te\'orica C-XI, Universidad Aut\'onoma de Madrid, 28049 Madrid, Spain }
\ead{luis.robledo@uam.es}
\author{P. Schuck}
\address{Institut de Physique Nucl\'eaire, CNRS, UMR8608,
  Orsay, F-91406, France}
\address{Universit\'e Paris-Sud, Orsay, F-91505, France    }
\address { Laboratoire de Physique et Mod\'elisation des Milieux Condens\'es, CNRS et
Universit\'e Joseph Fourier, 25 Av. des Martyrs, BP 166, F-38042 Grenoble Cedex 9, France  }
\ead{schuck@ipno.in2p3.fr}
\author{X. Vi\~nas}
\address{Departyament d'Estructura i constituents de la Mat\`eria
and Institut de Ci\`encies del Cosmos, Universitat de Barcelona,
 Av. Diagonal 647, E-08028 Barcelona, Spain    }
\ead{xavier@ecm.ub.es}
\begin{abstract}
In recent years impressive progress has been made in the development of highly accurate energy density
functionals, which allow to treat medium-heavy nuclei. In this approach one tries to describe not only the
ground state but also the first relevant excited states. In general, higher accuracy requires a larger set of
parameters, which must be carefully chosen to avoid redundancy. Following this line of development, it is
unavoidable that the connection of the functional with the bare nucleon-nucleon interaction becomes more and
more elusive. In principle, the construction of a density functional from a density matrix expansion based on
the effective nucleon-nucleon interaction is possible, and indeed the approach has been followed by few authors.
However, to what extent a density functional based on such a microscopic approach can reach the accuracy of the
fully phenomenological ones remains an open question. A related question is to establish which part of a
functional can be actually derived by a microscopic approach and which part, on the contrary, must be left as
purely phenomenological. In this paper we discuss the main problems that are encountered when the microscopic
approach is followed. To this purpose we will use the method we have recently introduced to illustrate the
different aspects of these problems. In particular we will discuss the possible connection of the density
functional with the nuclear matter Equation of State and the distinct features of finite size effects proper of
nuclei.
\end{abstract}
%\pacs{21.60.-n, 21.65.+f, 24.10.Cn, 31.15.Ew}

\maketitle

\section{Introduction}

The microscopic calculation of ground state energy and particle density of medium and heavy nuclei based on
realistic nucleon-nucleon interaction requires the solution of a formidable many-body problem. For this reason
effective nucleon-nucleon interactions have been introduced, like the Skyrme \cite{vauth} and Gogny \cite{gog}
forces. They simplify enormously the problem since, by construction, they must be used at the mean field level,
and the calculation  of the mean single particle potential and of the ground state energy becomes easily
manageable. The number of parameters which enter these effective forces is typically around ten and they are
adjusted to reproduce finite nuclei and some equilibrium nuclear matter properties. However, recently also data
from a theoretically determined neutron matter Equation of State (EOS) \cite{sly4} have been used as input (see
also an older attempt in this direction in \cite{Lamb}). Generally these forces give rise to an effective
nucleon mass $m^*<m$ with typically $m^*/m \simeq 0.7$ in the non-relativistic framework. With these ingredients
nuclear mean field theories are very successful to describe nuclear properties as, e.g. binding energies, radii.
Their use can be extended to the evaluation of more realistic single level schemes, of the nuclear excitation
spectra, in particular giant resonances, of the fission barriers, of the nucleon-nucleus optical potential, and
so on. The price to be payed is that the connection with the bare nucleon-nucleon interaction is not apparent.
The first attempt to connect the effective NN interaction and the underlying bare interaction was the density
matrix expansion of Negele and Vautherin \cite{NV72}, based on the assumption that the effective NN interaction
can be identified essentially with the Brueckner G-matrix calculated in nuclear matter and on the gradient
expansion of the non-local one-body density matrix. See also another attempts in the same direction in Ref.
\cite{Campi,Meyer,XavierDM}. This approach was recently generalized to include also the vector (spin) part of
the density matrix at the same level of accuracy as the scalar part \cite{Duguet2}. The challenging program of
deriving the effective interaction to be used in nuclear structure studies from the bare NN interaction has
still to be completed .\par
 An approach similar to the Skyrme-like one is the energy density functional (EDF) approach, where the basic
 quantity to start with is directly the functional that expresses the energy in terms of the matter
 density and its gradients. The functional must be minimized to obtain the actual ground state energy and matter
 density profile. Following the method of Kohn and Sham \cite{KS,jones,eschrig,Mat02,Per01,Per03,Per05},
 developed in atomic, molecular and
 solid state physics, the minimization procedure can be performed by introducing a set of auxiliary single particle
 states and taking for the kinetic energy the Slater form in this basis. In this way the minimization procedure
 gives Hartree-like equations for the single particle states, where the interaction part includes in an
 effective way the overall exchange and correlation contributions. The latter, in condensed matter physics, is taken from
 accurate calculations for the homogeneous electron system or in a purely phenomenological fashion.
This is not the usual way of proceeding in nuclear physics, although it exists, to our knowledge, an earlier
attempt in this direction, see ref. \cite{F9801}.
 \par
In this paper we will discuss to what extent the EDF approach in nuclear structure can be based on microscopic
many-body results on nuclear matter Equation of State (EOS) and which accuracy can be reached. To this purpose
we will rely on recent achievements along this line of research and we will analyze the main open problems that
hinder the development of the microscopic many-body theory of the nuclear EDF.

\section{Basic formalism}

 Despite the fact that the applicability of the KS-DFT approach to self-bound systems, as nuclei, is not
obvious \cite{Krei01,Eng07,Dob07,Gir07,workshop}, it is commonly believed that the basis of KS-DFT lies in the
Hohenberg-Kohn (HK) theorem
 \cite{HK}, which states that for a Fermi system, with a non-degenerate
ground state, the total energy can be expressed as a functional of the density $\rho({\bf r})$ only. Such a
functional reaches its variational minimum when evaluated with the exact ground state density.  Furthermore, in
the standard KS-DFT method one introduces an auxiliary set of $A$ orthonormal single particle wave functions
$\psi_i({\bf r} )$, where $A$ is the number of particles, and the density is assumed to be given by
\begin{equation}
 \rho( {\bf r} ) \, =\, \Sigma_{i,s,t} | \psi_i( {\bf r},s,t ) |^2
\label{e:eq1} \end{equation} \noindent where $s$ and $t$ stand for spin and iso-spin indices. The variational
procedure to minimize the functional is performed in terms of the orbitals instead of the density. As in
condensed matter and atomic physics the HK functional $E[\rho({\bf r})$] is split into two parts: $E =
T_0[\rho]+W[\rho]$ \cite{KS}. The first piece $T_0$ corresponds to  the uncorrelated part of the kinetic energy
and within the KS method it is written as
\begin{equation}
 T_0=\frac{{\hbar}^2}{2m} \sum_{i,s,t} \int
d^3r|\nabla \psi_i( {\bf r},s,t ) |^2. \label{eq:eq2} \end{equation} \noindent The other piece $W[\rho]$
contains the potential energy as well as the correlated part of the kinetic energy.

 Then, upon variation, one gets a closed
set of $A$ Hartree-like equations with an effective potential, the functional derivative of $W[\rho]$ with
respect to the local density $\rho({\bf r})$. Since the latter depends on the density, and therefore on the
 $\psi_i$'s, a self-consistent procedure is
necessary. The equations are exact but they only can be of some use if a reliable approximation is found for the
otherwise unknown density functional $W[\rho]$. It has to be stressed that in the KS-DFT formalism the exact
ground state wave function is actually not known, the density being the basic quantity. In nuclear physics,
contrary to the situation in condensed matter and atomic physics, the contribution of the spin-orbit interaction
to the energy functional is very important.
 Non-local contributions have been included in DFT in several ways already
long ago (see \cite{Eng03} for a recent review of this topic). Consequently, the spin-orbit part also can be
split in an uncorrelated part $E^{s.o.}$ plus a remainder. The form of the uncorrelated spin-orbit part is taken
exactly as in the Skyrme \cite{vauth} or Gogny forces \cite{gog}.
 We thus
write for the functional in the nuclear case $E = T_0 + E^{s.o.} + E_{int} + E_C$, where we explicitly split off
the Coulomb energy $E_C$ because it is a quite distinct part in the Hamiltonian. It shall be treated, as usual,
at lowest order, i.e. the direct term plus the exchange contribution in the Slater approximation, that is
$E_C^H= (1/2) \int \int d^3rd^3r' \rho_p({\bf r})|{\bf r}-{\bf r'}|^{-1} \rho_p({\bf r'})$,
 and
$E_C^{ex} = -(3/4)(3/\pi)^{1/3} \int d^3r {\rho_p({\bf r})}^{4/3}$ with $E_C=E_C^{H} + E_C^{ex}$ and
$\rho_{p/n}$ the proton/neutron density.
\par Let us now discuss the nuclear energy functional
contribution $E_ {int}[\rho_n,\rho_p]$ which contains the nuclear potential energy as well as additional
correlations. We shall split it in a finite range term $E_ {int}^{FR}[\rho_n,\rho_p]$ to account for correct
surface properties and a bulk correlation part $E_ {int}^{\infty}[\rho_n,\rho_p]$ that we take from a
microscopic infinite nuclear matter calculation \cite{BMSV}. Thus our final KS-DFT -like functional reads:
\begin{equation}
E = T_0 + E^{s.o.} + E_{int}^{\infty} + E_{int}^{FR} + E_C. \label{eq:eq6} \end{equation}

For the finite range term we make the simplest phenomenological ansatz possible
\begin{eqnarray}
E_{int}^{FR}[\rho_n,\rho_p ] &=& \frac{1}{2}\sum_{t,t'}\int \int d^3r d^3r'\rho_{t}({\bf r}) v_{t,t'}({\bf
r}-{\bf r'})\rho_{t'}({\bf r'} )
\nonumber \\
&-& \frac{1}{2}\sum_{t,t'} \gamma_{t,t'}\int d^3r {\rho_{t}({\bf r})} \rho_{t'}({\bf r}) \label{eq:eq7}
\end{eqnarray} with $t=$ proton/neutron and $\gamma_{t,t'}$ the volume integral of $v_{t,t'}(r)$. The
substraction in (\ref{eq:eq7}) is made in order not to contaminate the bulk part, determined from the
microscopic infinite matter calculation. Finite range terms have already been used earlier, generalizing usual
Skyrme functionals (see e.g \cite{BKN,Umar,Fay00}). In this study, for the finite range form factor
$v_{t,t'}(r)$ we make a simple Gaussian ansatz: $v_{t,t'}(r)=V_{t,t'}e^{-r^2/{r_0}^2}$. We choose a minimum of
three open parameters: $V_{p,p}=V_{n,n}=V_L, V_{n,p}=V_{p,n}=V_U$, and $r_0$.
 The only undetermined and most important
piece in (\ref{eq:eq6}) is then the bulk contribution $E_{int}^{\infty}$. As already mentioned, we obtain
$E_{int}^{\infty}$ from microscopic infinite matter calculations, using a realistic bare force, together with a
converged hole-line expansion \cite{BMSV}. We first reproduce by interpolating functions the {\it correlation}
part of the ground state energy per particle of symmetric and pure neutron matters, and then make a quadratic
interpolation for asymmetric matter.
 Finally the total correlation
contribution to the energy functional in local density approximation reads:
\begin{equation}
E_{int}^{\infty}[\rho_p, \rho_n] = \int d^3r \big[ P_s(\rho) (1 - \beta^2) + P_n (\rho)\beta^2 \big] \rho
\label{eq:eq4a}
\end{equation}
 where $P_s$ and $P_n$ are two interpolating polynomials
for symmetric and pure neutron matter, respectively, at the density $\rho = \rho_p + \rho_n$, and $\beta =
(\rho_n - \rho_p)/\rho$ is the asymmetry parameter. The interpolating polynomial for symmetric matter has been
constrained to allow a minimum exactly at the energy $E/A$ = - 16.00 MeV and Fermi momentum $k_F$ = 1.36
fm$^{-1}$, i.e. $\rho_0$= 0.16 fm$^{-3}$. This is within the uncertainty of the numerical microscopic
calculations of the EOS. It has to be stressed that the use of a polynomial in density is just for practical
reasons.

The constrained fit was performed by keeping the EOS as smooth as possible, thus allowing for some very small
deviations from the microscopic calculations below saturation density. An interpolating fit which goes exactly
through the calculated EOS, as performed in \cite{BMSV}, gives a not good enough saturation point (typically E/A
= -15.6 MeV, $k_F$=1.38 fm$^{-1}$).

 As discussed in \cite{Mar07}, the low density behavior of the nuclear matter
EOS is quite intricate and usually not reproduced by Skyrme and Gogny functionals ( see also ref. \cite{BMSV}),
missing quite a substantial part of binding. We show our EOS for nuclear and neutron matter in Fig. 1. Since we
want to construct the EDF, as much as possible, on the basis of the microscopic calculations, the bulk part
$E_{int}^{\infty}$ of the functional, directly related to bare NN and TBF (three-body forces), is determined
once and for all and we will use it in (3) together with LDA.

\begin{figure}
\begin{center}
\includegraphics[width=15.cm]{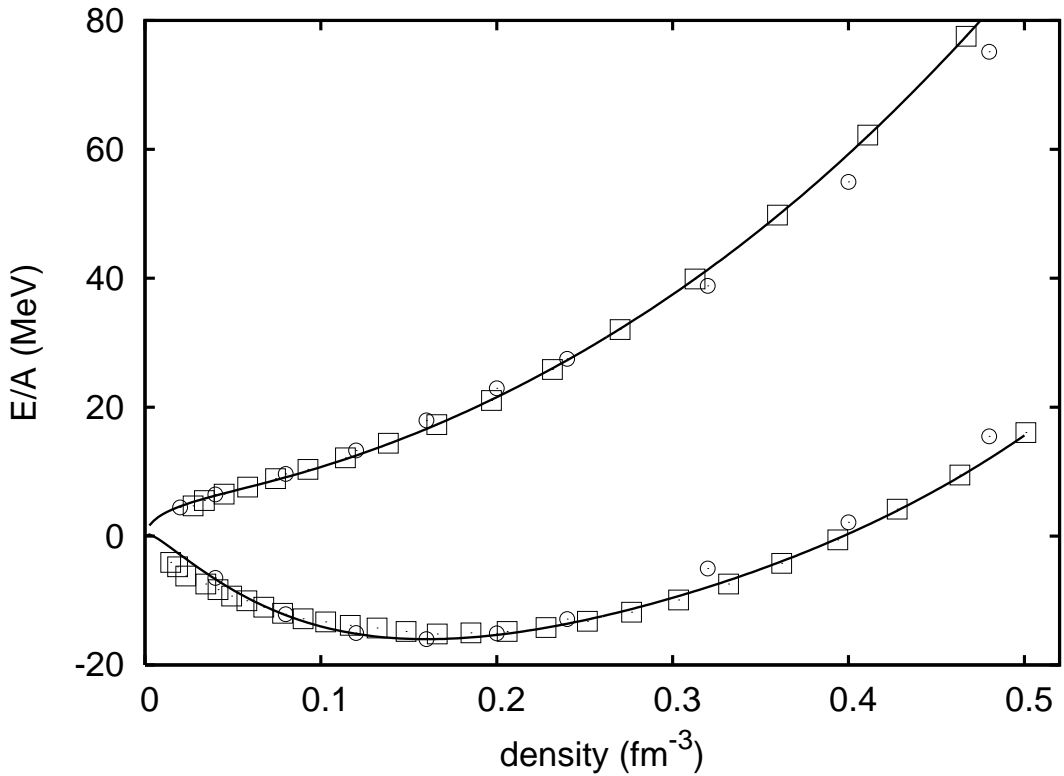}
\caption{EOS of symmetric and neutron matter obtained by the microscopic calculation (squares) and the
corresponding polynomial fits (solid lines).For comparison the microscopic EOS of Refs. \cite{AP} are also
displayed by open circles}
\end{center}
\label{fig:EOS}
\end{figure}

The only open parameters are, therefore, the ones contained in the finite range surface part, Eq.(\ref{eq:eq7}),
and the strength of the spin-orbit contribution, that we fit to reproduce finite nuclei properties. We, thus,
follow {\it exactly} the strategy employed in condensed matter. On the contrary, in nuclear physics, almost
exclusively a different strategy is usually adopted (see, however, Ref.\cite{F9801} with some ingredients
similar as in the present approach ): functionals like the one of (\ref{eq:eq6}), i.e. bulk, surface, etc., were
globally parametrized with typically of the order of ten parameters which, then, were determined fitting {\it
simultaneously} some equilibrium nuclear matter (binding energy per particle, saturation density,
incompressibiliy, etc.) and finite nuclei properties. However, in this way, bulk and surface are not properly
separated and early attempts used to miss important infinite nuclear matter properties, as, e.g. stability of
neutron matter at high density \cite{gog} and other stability criteria. Modern Skyrme forces , like the
Saclay-Lyon (SLy) ones, explicitly use the high density part ($\rho/\rho_0 > 0.65$) of microscopic neutron
matter calculations for the EOS in the fitting procedure \cite{sly4} and thus avoid collapse. Therefore, modern
functionals usually reproduce reasonably well microscopically determined EOS for neutron and nuclear matter
\cite{BMSV}. Examples are, among others,
 the
SLy-forces \cite{sly4} (see Fig.1 in \cite {BMSV}) and the Fayans functional DF3 (see e.g. Fig. 3 in
\cite{Fay00}). In this presentation we follow the just discussed alternative approach, that is different on a
qualitative level from the usual and allows, via the fit, to reproduce very accurately the microscopic infinite
matter results in the whole range of densities considered. This may be important for surface properties and
neutron skins in exotic nuclei, what shall be investigated in the future.

For open shell nuclei, we still have to add  pairing. The formal generalization of the rigorous HK theorem to
paired system has been given in Ref. \cite{Oli88}. In  the present  work our main objective is to discuss the
KS-DFT scheme for the non pairing part, thus we add pairing in a very simple way within the BCS approach. For
this we simply take the density dependent delta force defined in Ref \cite{GSGS} for $m=m^*$  with the same
parameters and in particular with the same cutoff. As far as this amounts to a cutoff of $\sim$ 10 MeV into the
continuum for finite nuclei, we have to deal with single-particle energy levels lying in the continuum. We have
simulated it by taking in the pairing window all the quasi-bound levels, i.e. the levels retained by the
centrifugal (neutrons) and centrifugal plus Coulomb (protons) barriers. This treatment of the continuum works
properly, at least for nuclei not far from the stability valley as it has been extensively shown in
\cite{Estal}. In this way we obtain two-neutron ($S_{2n}$) and two-proton ($S_{2p}$) energy separations for
magic proton and neutron numbers in quite good agreement with the experiment (see also below).

In our calculations the two-body center of mass correction has been included in the self-consistent calculation
using the pocket formula, based on the harmonic oscillator, derived in Ref.\cite{BSM} which nicely reproduces
the exact correction as it has been shown in \cite{STV}. Our functional is now fully defined and, henceforth, we
call it BCP-functional. Preliminary results based on this functional have been presented in ref.
\cite{BCP,Robledo.08}.

\section{Fitting procedure and general results}

In this section we discuss the fitting protocol to determine the open parameters of our functional and the
accuracy that can be reached following the scheme described in the previous section. To this purpose we make a
comparison between our recent results and the ones that can be obtained with the Gogny functional, which is one
of the most accurate phenomenological functional. In this paper we follow a novel strategy, recently proposed
\cite{Niksic.08}, for the fitting procedure to obtain the set of parameters for the surface and spin-orbit part
of the functional, that we call the BCP09 functional. Contrary to our previous works we no longer use the
binding energies of spherical nuclei for the fit but, following the suggestion of Ref. \cite{Niksic.08}, we use
a set of deformed nuclei carefully chosen in the rare earth, actinide and super-heavy regions instead. The
underlying philosophy behind this choice \cite{Niksic.08} is that for spherical nuclei the fluctuations in the
two most relevant collective variables, namely, the quadrupole deformation and the pairing correlations are
large (i.e. the minimum in those variables is broad in some sense) and therefore correlations beyond mean field
are important and difficult to evaluate. On the other hand, for deformed nuclei the minimum as a function of the
quadrupole degree of freedom is stiffer than its  counterpart in spherical nuclei and the additional
correlations (rotational energy correction mainly) are not so difficult to compute in a mean field model.

In order to fit the three free parameters of the surface part ($V_U$, $V_L$ and $r_0$) of BCP and the spin-orbit
strength $W_{\textrm{LS}}$ we have taken as in \cite{Niksic.08} 84 well deformed nuclei in the rare earth,
actinide and super-heavy regions where the experimental binding energies are known. The four parameters are
determined by minimizing the mean square root deviation of the binding energy $\sigma_E$. The theoretical
binding energy has been computed as the Hartree- Fock- Bogoliubov (HFB) mean field energy plus a rotational
energy correction, which differs from the standard one \cite{RS.80} by a phenomenological factor to account for
the approximate calculation of the Yoccoz moment of inertia. We also consider an additional correction factor
(not relevant in the present calculation) to deal with the weak deformation regime  (see, for instance,
\cite{Egido.03}). Also a correction to the binding energy ensuing from the finite size of the harmonic
oscillator basis used, as estimated in \cite{Hilaire.07}, has been included. Axial symmetry has been preserved
in the HFB calculation as the nuclei chosen are not expected to develop triaxiality. On the other hand
reflection symmetry is allowed to be broken in the solution of the HFB equation, to allow for octupole
deformation in the ground state which is relevant for a few of the actinides considered. The calculation of the
84 ground states for a set of parameters can be carried out in a powerful personal computer in around half an
hour. As a consequence of this figure, an unconstrained and blind search of the four parameters leading to the
absolute minimum of $ \sigma_E$ is a task out of the scope of the present exploratory considerations.
Fortunately, the expression for the binding energy depends linearly on three  out of the four free parameters
($V_U$, $V_L$ and $W_{\textrm{LS}}$) and the method suggested by Bertsch et al. \cite{Bertsch.05} in those
situation should work well. We have implemented Bertsch's procedure for the three parameters $V_U$, $V_L$ and
$W_{\textrm{LS}}$ and performed systematic calculations for as a function of the other parameter $r_0$. Usually,
for a given $r_0$ value and with a reasonable choice of starting parameters, the linearized Bertsch's method
leads to a local minimum of $\sigma_E$ in a couple of iterations. The reason is that the correlation matrix of
Bertsch's method has in our case two eigenvalues which are two  to three orders of magnitude smaller than the
remaining one. This fact points to the existence of only  one free parameter (a linear combination of $V_U$,
$V_L$ and $W_{\textrm{LS}}$ with weights corresponding to the components of the eigenvector of the largest
eigenvalue).

At this point it has to be mentioned that the fact that the expression of the binding energy is linear in $V_U$,
$V_L$ and $W_{\textrm{LS}}$ by no means imply the value of binding energy to be globally a linear function of
those quantities. The reason is that the binding energy is obtained after a self-consistent HFB calculation and
the self-consistency breaks the linear dependence of the wave function on the parameters mentioned above. As a
consequence, Bertsch method is only valid locally what implies that  the procedure will end up in different
minimum depending upon the values of the initial parameters. A careful analysis of the results obtained so far
indicates that the spin-orbit strength is the most sensitive starting parameter and therefore a search on this
degree of freedom should be carried out in addition to the search on $r_0$.

Using this new protocol we have performed a fit using for the bulk density dependent part of the functional new
interpolating polynomials for the nuclear matter equation of state in a wider range of density with respect to
the previous ones (BCP1 and BCP2 in refs. \cite{BCP,Robledo.08}).
The new fit has still a binding energy per nucleon $E/A=16$ MeV at saturation density $\rho_0=0.16$ fm $^{-3}$
(for a compressibility of 220 MeV). For the spin-orbit strength we choose to stay around the $W_\textrm{LS}=95$
MeV value and perform a minimization of $\sigma_E$ for different values of $r_0$ from 0.85 fm up to 1.00 fm. The
minimum value of $\sigma_E$ was obtained for $r_0=0.9$ fm. The values of the other parameters are $V_U=
-137.024$ MeV, $ -117.854$ MeV and $W_\textrm{LS}=  95.43$ MeV fm$^5$ for a value of $\sigma_E$ of 0.545 MeV. In
order to check for the suitability of the spin-orbit strength value used we performed another two minimizations
of $\sigma_E$ fixing $r_0$ at 0.9 fm but taking $W_\textrm{LS}= 90.90$ MeV fm$^5$ and 102.51 MeV fm$^5$. The
values of $\sigma_E$ obtained are 0.569 MeV and 0.566 MeV respectively, supporting the assignment of
$W_\textrm{LS}= 95.43 $ MeV fm$^5$ as the value leading to the lowest $\sigma_E$ (this check is incomplete as a
more extensive search in the two parameters should be performed; work in this direction is in progress and will
be reported in a near future).

\begin{figure}
\begin{center}
\includegraphics[angle=270,width=0.85\columnwidth]{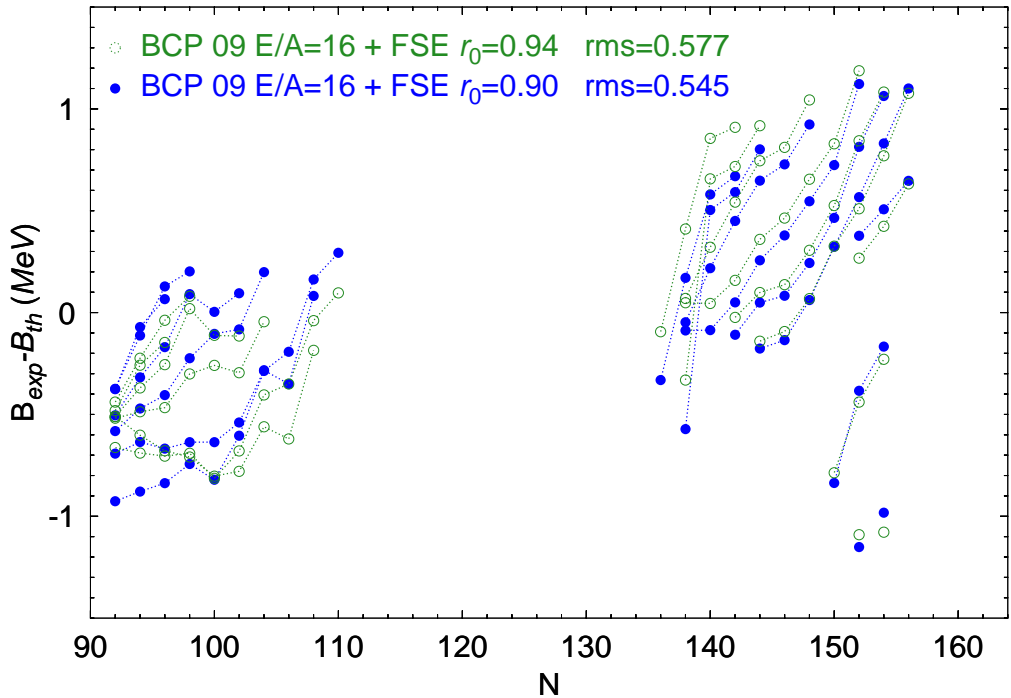}
\end{center}
\caption{(Color online) The difference in binding energies with respect to the experimental value for the 84
nuclei considered as a function of neutron number. Results for $E/A=$16 MeV, $r_0=0.9$ fm that yield
$\sigma_E=0.545$ MeV are plotted as full (blue) dots, whereas results for $E/A=$16 MeV, $r_0=0.94$ fm that yield
$\sigma_E=0.577$ MeV are plotted as open (green) circles. The group of points to the left corresponds to the
rare earth isotopes whereas the one to the right are for the actinides and superheavy.\label{fig:SigmaBCP09}}

\end{figure}

In Fig. \ref{fig:SigmaBCP09} we show the individual differences in binding energy ($B$) between the theoretical
results and the experimental ones for the 84 nuclei considered. The quantity $\Delta
B=B_\textrm{Exp}-B_\textrm{Th}$ has been plotted as a function of the number of neutrons $N$ and the values
corresponding to the same isotope are joined with lines. Two sets of parameters have been considered, namely one
with $r_0=0.9$ fm producing the lowest value of $\sigma_E$ (full dots) and other with a slightly higher value
$\sigma_E=0.577$ MeV and obtained fixing $r_0=0.94$ fm (circles). It is worth pointing out that the absolute
value of $\Delta B$ never exceeds 1.2 MeV. We also observe that the change in parameters only produce minor
changes in the $\Delta B$'s. Finally, we notice that the actinides and superheavies look a little bit too
underbound as compared to the rare earth isotopes. Finding the origin of such a relative underbinding will help
to improve the quality of the fit reducing the value of $\sigma_E$. Although the value of $\sigma_E$ was
obtained by considering an optimal set of deformed nuclei ( those with a well established deformation ) it is
very likely that the $\sigma_E$ value for a complete set of nuclei from proton drip line to neutron drip line
will be promising.

\begin{figure}
\begin{center}
\includegraphics[angle=270,width=0.85\columnwidth]{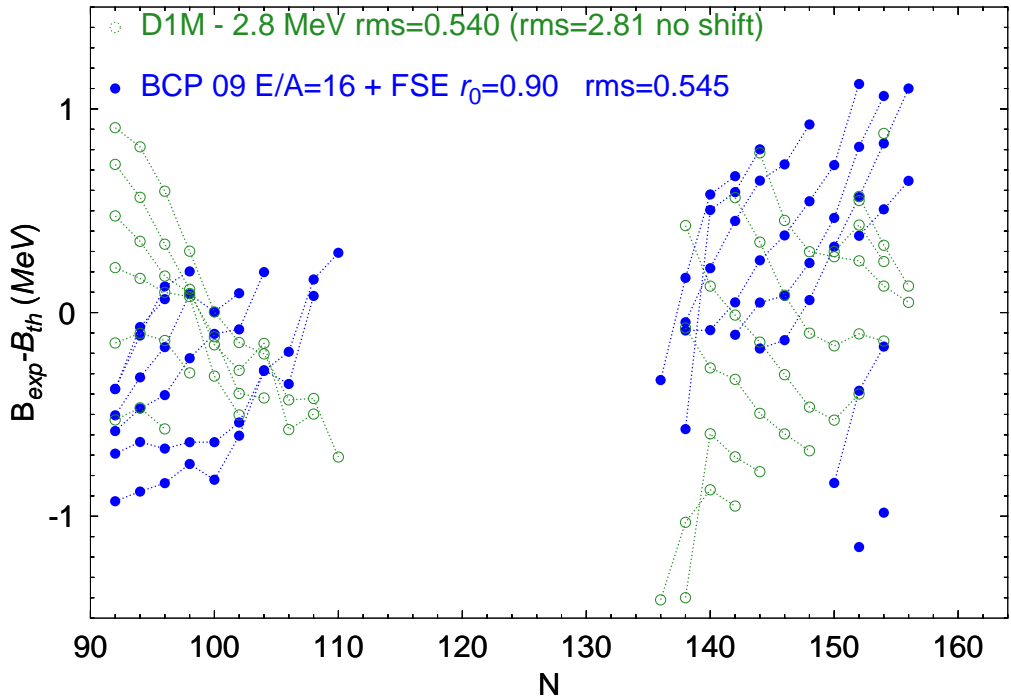}
\end{center}
\caption{(Color online) Same as Fig. \ref{fig:SigmaBCP09} but this time the BCP results with $E/A=$16 MeV,
$r_0=0.9$ fm that yield $\sigma_E=0.545$ MeV are compared to those of the Gogny D1M force. See text for further
details. \label{fig:SigmaBCP09D1M}}

\end{figure}

In order to compare the results obtained with BCP 09 with others of other interactions we have performed
calculations with the Gogny force and the most recent parametrization D1M that was made up with the idea of
producing a mass table. Unfortunately, the details of how the theoretical binding energy was obtained are not
detailed enough as to allow a reproduction of them (the main uncertainty is in the quadrupole motion zero point
energy correction). We have decided to use the same protocol as with BCP 09, that is, the rotational energy
correction and the finite size of the basis effect are included, and we have left out the zero point energy
correction mentioned above. As a consequence, our values for $\sigma_E$ of D1M are too high and in order to make
a more fair comparison we just shifted all the binding energies computed by a constant quantity (2.8 MeV)
obtained as to minimize $\sigma_E$. With this readjustment we obtain for D1M and the 84 nuclei considered the
value $\sigma_E=0.54$ MeV which is slightly better than our value but, on the other hand, of similar quality to
the one of the BCP 09 functional. In Fig. \ref{fig:SigmaBCP09D1M} we have plotted the individual values of
$\Delta B$ both for BCP 09 and Gogny D1M. We notice how the behavior of $\Delta B$ for the Gogny D1M as a
function of $N$ for each isotopic chain is different from the one of BCP 09; decreasing with increasing $N$ in
the case of Gogny D1M whereas it is increasing with increasing $N$ for BCP 09. It is also worth to point out
that Gogny D1M force produces results in the actinide and super-heavy nuclei which are more spread out than the
ones of the BCP 09 functional. By comparing both results we can conclude that, at least for the 84 deformed
nuclei considered, the performance of BCP 09 is as good as the one of Gogny D1M and at a fraction of the
computational cost (the absence of exchange terms and non local pairing in BCP makes the numerical evaluation of
the quantities entering the HFB equation much faster  than for the Gogny D1M case). This is an encouraging
starting point for a more detailed study of the viability of BCP 09 in the description of low energy nuclear
structure.

Next we want to explore the ability of our proposed BCP09 energy density functional to describe nuclear
ground-state properties of other nuclei not included in the fitting protocol used to obtain  the free parameters
that describe the surface and spin-orbit contributions to the functional.

\begin{figure}
\includegraphics[width=12cm,angle=-90]{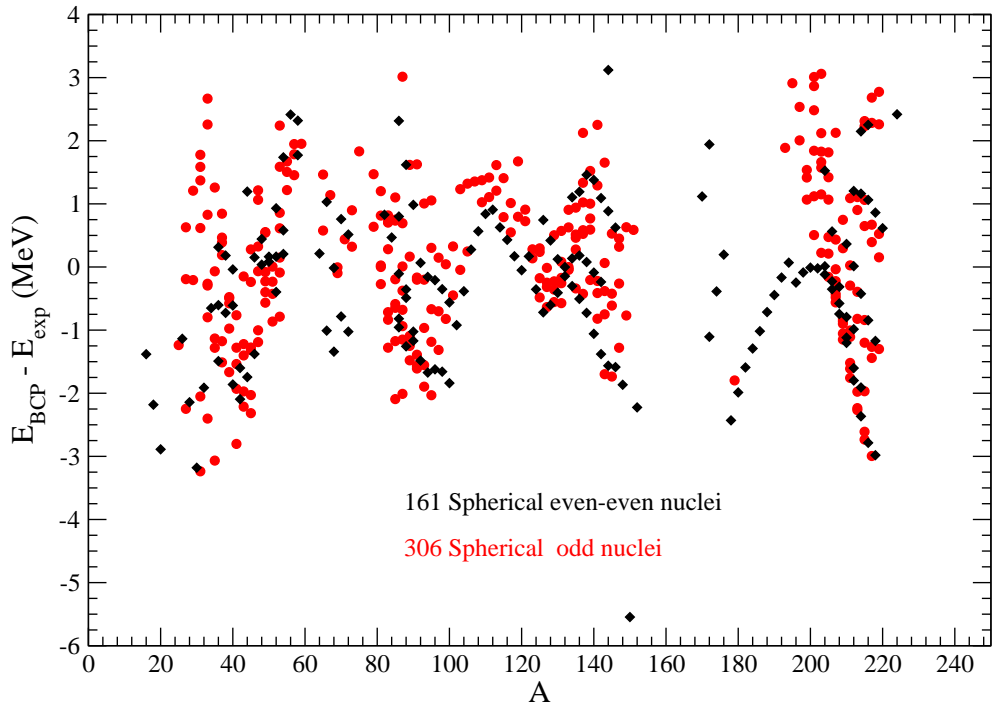}
\caption{(Color online) Energy differences as a function of mass number for a set of 161 spherical nuclei (black
points) and 306 spherical odd nuclei (red points). \label{fig:DiffBSph}}
\end{figure}

To this end we have computed the ground-state energies of 161 even-even and 306 odd spherical nuclei. These
nuclei are chosen to be spherical according to the deformation properties of the compilation of M\"oller and Nix
\cite{Moller.97}. To deal with odd nuclei, we have used the blocking approach on top of the BCS calculation. The
differences $\Delta B = E_\textrm{BCP} - E_\textrm{Exp}$, where $E_\textrm{BCP}$ are the theoretical predictions
of the BCP09 functional and $E_\textrm{Exp}$ the experimental values taken from \cite{Audi.03}, are displayed in
Fig. \ref{fig:DiffBSph}. The agreement found between the theoretical prediction and the experimental values is
fairly good finding an energy rms $\sigma_E \simeq$ 1.3 MeV. It should be mention that both, even-even and odd
nuclei, give basically the same rms when they are considered separately. From the figure we can see that the
theoretical ground-state energies are scattered, rather symmetrically, around $\Delta B=0$ within a window of
about $\pm$ 3 MeV with the only exception of the nucleus $^{150}_{68}$Er. Combining these values for 467
spherical nuclei with the results obtained for the 84 deformed nuclei used in the fitting protocol, we find a
global rms $\sigma_E \simeq $1.2 MeV, a value that can be improved if the analysis of the ground-state energies
include more deformed nuclei.

\begin{figure}
\includegraphics[width=12cm,angle=-90]{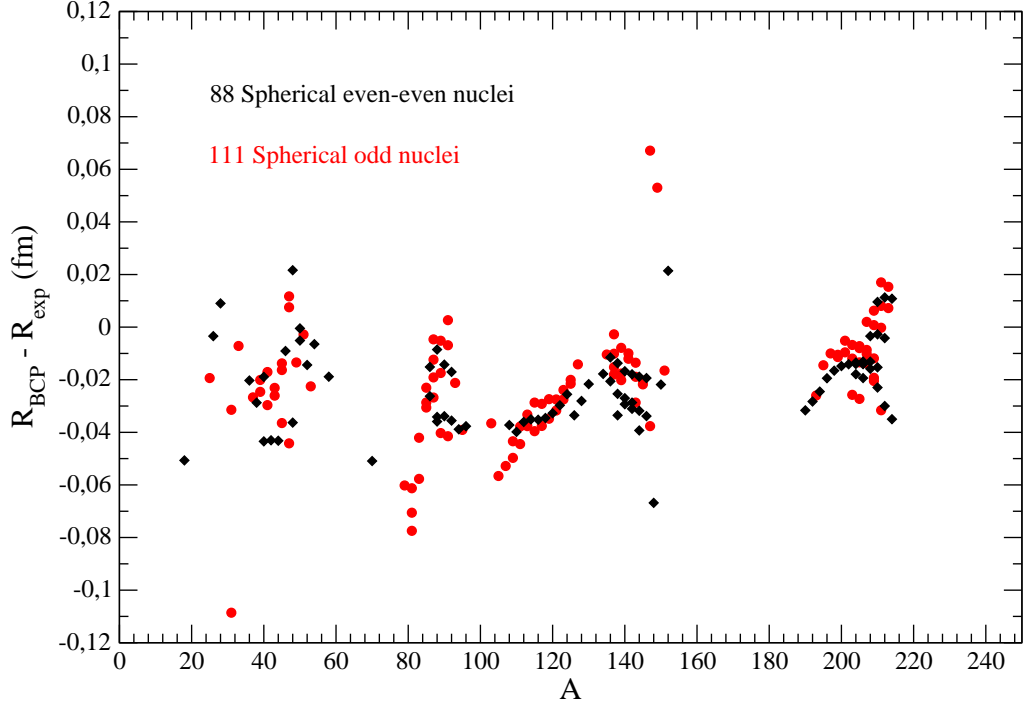}
\caption{(Color online) Differences of radii are shown as a function of mass number for a set of 88 spherical
nuclei (black points) and 111 spherical odd nuclei (red points)\label{fig:DiffRSph}}
\end{figure}

In Fig. \ref{fig:DiffRSph} we show the differences between the calculated and experimental charge radii $\Delta
R = R_\textrm{BCP} - R_\textrm{Exp}$. In this Figure we have displayed  the differences $\Delta R$ for 88
even-even and 111 odd spherical nuclei for which the experimental charge radii are known \cite{Angeli.04}. From
this Figure we see that the BCP09 energy density functional predicts, on the average, smaller charge radii than
the experimental values. With few exceptions, the theoretical charge radii are scattered within a window of 0.04
fm around an average values $\Delta R$ = -0.02 fm. The global quality of this fit of the charge radii is given
by rms $\sigma_R$=0.030 fm and again the rms of the even-even and odd nuclei are basically the same. This global
rms can be compared with one provided by the HFB-8 model \cite{Buchinger.05} which is $\sigma_R$=0.0275 fm,
however computed using 782 experimental data. As compared with the results provided by the earlier BCP2
functional, that included some experimental charge radii in the fit, the $\sigma_R$ values are roughly the same,
but in the case of the BCP2 functional the theoretical charge radii are rather overestimated, at least for light
nuclei.

\begin{figure}
\begin{center}
\includegraphics[width=0.85\columnwidth]{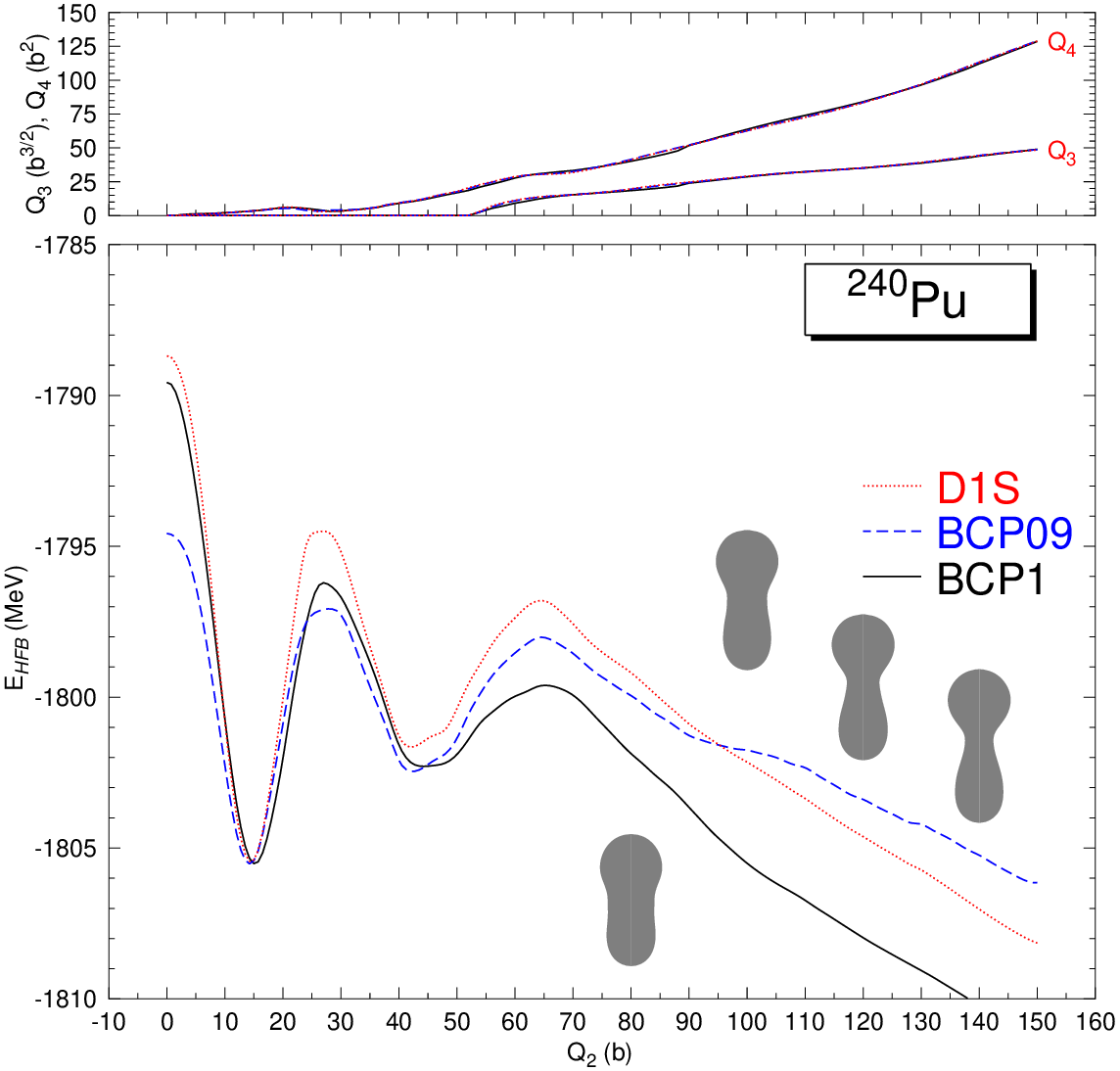}
\end{center}
\caption{(Color online) In the lower panel the potential energy curve as a function of the quadrupole moment
$Q_2$ for the two BCP functionals and Gogny force used in the calculations is shown. In the upper panel, the
octupole and hexadecapole moments are displayed again for the three interactions/force used. The lines lie on
top of each other and are hardly distinguisable. See text for further details. \label{fig:MFProp240Pu}}

\end{figure}

After having defined a performing functional in the realm of nuclear masses one has to check its suitability in
describing other nuclear properties like deformations. A detailed study of deformations with BCP 09 is underway
and will be reported in the future, but in order to give a taste of the results already obtained we show in Fig.
\ref{fig:MFProp240Pu} the typical outcome of a fission barrier calculation for the nucleus $^{240}$Pu that has
also been discussed in Ref. \cite{Robledo.08} for early versions of the BCP functional. In fission studies the
energy of the system obtained in the HFB framework is obtained as a function of the quadrupole moment by
constraining the HFB solution. In this way a potential energy curve (PEC) is obtained that in the present case
is displayed in the lower panel of Fig. \ref{fig:MFProp240Pu} for the BCP1 and BCP09 functionals and the Gogny
D1S force. The BCP1 and BCP09 curves have been shifted by 0.65 MeV and 4.5 MeV respectively as to make the first
minimum at $Q_2=14 $ b ($\beta_2=$ 0.26 ) coincide in energy. After the first (ground state) minimum there is a
barrier (first fission barrier) that is higher for D1S than for the BCP functionals. The height of this barrier
is known to be very sensitive to triaxial effects which are still not accounted for in the present calculation.
After the first fission barrier there is an additional excited local minimum at $Q_2=46$ b ($\beta_2=$0.73) that
corresponds to the fission isomer. The excitation energy of the fission isomer obtained with the two BCP
functionals is more or less the same with a value of 3.5 MeV. For the Gogny D1S the result is slightly higher
(around 4.5 MeV). The following maximum corresponds to the second fission barrier which is definitely much
higher for Gogny D1S than for the BCP functional. After the second fission barrier the typical decrease of the
energy consequence of the Coulomb repulsion between the nascent fragments (the real shape of the nucleus is
depicted at the corresponding $Q_2$ value as a contour plot of the real density at a value of the contour of
0.08 fm$^{-3}$). As the two fragments are still connected by the neck the nuclear part of the interaction still
has an effect, explaining the differences observed between the three calculations. Finally, in the upper panel
of Fig. \ref{fig:MFProp240Pu} the octupole ($Q_3$) and hexadecapole ($Q_4$) moments of the mass distribution
obtained with the three functionals/interactions are shown. The values lie on top of each others indicating (as
the multipole moments considered are the most relevant) that the shapes of the nucleus along the whole fission
path are essentially the same for the three calculations.

\section{Open problems}

The problems to be clarified in the EDF approach can be in general classified in two categories. On one hand,
the use of the nuclear matter EOS must be taken with care since the microscopic calculations cannot reach the
accuracy usually required in nuclear structure calculations and contains intrinsic and unavoidable uncertainties
discussed below. On the other hand, finite size effects must be included in the functional with a minimal set of
parameters. The procedure of doing that is not straightforward and requires extensive analysis.

\subsection{Nuclear matter EOS}

Following the KS scheme, the nuclear EOS calculated microscopically should be included in the EDF mainly by an
analytic fit that reproduces accurately both the nuclear matter and neutron matter results. In this way the
saturation point, as well the whole density dependence of the binding energy, can be incorporated in the EDF.
However, the microscopic calculations cannot be considered without any uncertainty. Besides the point of
numerical accuracy, it is well known that it is necessary to introduce three-body forces in the microscopic
calculations in order to reproduce the correct saturation point, which however contains necessarily a
phenomenological uncertainty. Therefore the TBF can be adjusted to reproduce different saturation points.
Furthermore the microscopic data do not fix uniquely the fitted EOS. We found that the quality of the nuclear
data fitting is very sensitive to the saturation point of the interpolated EOS. The optimal value of the energy
per particle at saturation seems to be close to -16 MeV, but even a shift as small as 0.1 MeV of this quantity
can deteriorate appreciably the quality of the nuclear data fit. This should be a general feature, independent
of the particular EDF that is used. Less sensitivity seems to be present for the saturation density. For the EOS
obtained from the BHF theory, the TBF contribution below saturation is quite small and the BHF is expected to be
accurate at low density, so that the main problem can be considered as how to fix the saturation point.\par Of
course the just discussed sensitivity on the saturation point could be considered a not serious problem, in view
of the phenomenological uncertainty that affects its position.\par All these problems appear absent for the
neutron matter EOS, for which the TBF contribution is smaller and correlation energy is substantially
reduced.\par

A separate discussion is needed for the type and strength of the TBF that are used in microscopic nuclear matter
calculations. The numerical calculations of the binding energy of nuclear few-body systems, like triton and
alpha particle, can be performed numerically very accurately. If realistic two-body forces, like the Argonne
v$_{18}$ potential, are used the binding energies turn out systematically underestimated. To remedy to this
drawback usually one adds a semi-phenomenological TBF, like the Urbana model \cite{Urbana}, that is adjusted to
reproduce the experimental binding energies. The unpleasant discovery is that the use of the same TBF in
microscopic calculations for nuclear matter does not produce a good saturation point. The TBF appear too
repulsive. Indeed, in general semi-phenomenological TBF's  contain a repulsive and attractive part, and the
standard procedure is to reduce the repulsive component in order to get a good saturation point. At
phenomenological level this means that one needs at least a four-body force in a non-relativistic many-body
scheme. \par It has to be mentioned in this respect that non-local two-body forces \cite{Dole} which are able to
reproduce the binding energy of few-nucleons systems fail to reproduce the correct saturation point, that then
turns out to be at too high density and too low energy \cite{nonloc}.
\par
Recently \cite{Bogner1} it has been shown that one can get a reasonable saturation point by keeping the TBF
obtained by fitting few-body systems if for the effective two-body force the so-called V$_{low}$ interaction is
used. The latter is obtained by projecting out the high momenta of the bare NN interaction and renormalizing
accordingly the interaction at low momenta. The resulting interaction is then phase-equivalent to the original
bare NN interaction up to the momentum cutoff, but it is much softer, and therefore it can be treated
perturbatively. Nuclear matter calculations with V$_{low}$ are not saturating, most probably because they misses
the effects of the Pauli principle and of the self-consistent single particle potential, which are distint
features  of the G-matrix. The final saturation point is a consequence of the compensation between the too large
attraction of V$_{low}$ and the too large repulsion of the TBF. All these uncertainties are of course embodied
in the final EOS and affect the detailed properties of the resulting EDF.

 Unfortunately there is a more basic problem for the low density EOS. At low enough density symmetric matter
 is expected to be unstable towards cluster formation. The picture of an homogeneous matter cannot be kept at
 very low density. The real EOS is therefore altered by the appearance of clusters of different sizes. However the EOS
 which includes cluster formation cannot be used in the EDF, since such a proliferation of light nuclei cannot
 be present in the low density region of finite nuclei. This is a general problem for any EDF.
The absence of clusters at the nuclear surface is due to the small value of the diffusion of the density
profile, which prevents any long range correlation to dominate. This fact offers a partial justification of
taking the homogeneous matter EOS at the BHF level. Indeed, short range correlations are well taken into account
by the BHF procedure, and therefore the corresponding EOS should be able to describe the properties of the bulk
contribution to the nuclear EDF even in the surface region. \par This problem still needs further analysis
before it can be considered satisfactorily solved.

\subsection{Incorporating the nuclear matter EOS in the EDF}

In the standard KS procedure the kinetic energy contribution is taken at the independent particle level with an
effective mass equal to the bare one. This is in agreement with the BHF scheme, where the kinetic energy part is
also kept at the free value and the whole correlation contribution is included in the interaction energy part
coming from the G-matrix. This means that also the correlated part of the kinetic energy is included in the
G-matrix contribution. Indeed the momentum dependence of the G-matrix is clearly the origin of the deviation of
the effective mass value from the bare one. It would not be difficult to separate this effect from the G-matrix
contribution, so that the kinetic energy would include a (density dependent) effective mass, while leaving the
rest of correlations as a genuine interaction energy. This leaves some freedom to the way the nuclear matter EOS
is actually included in the EDF, which allows to go beyond the usual KS scheme. It can be expected that
different procedures are not necessarily equivalent. This problem has still to be analyzed in detail.

\subsection{Finite size effects}

The density matrix expansion of Negele and Vautherin \cite{NV72} suggests that the surface contribution to the
binding energy of nuclei may also be extracted from nuclear matter calculations. This would be very welcome,
since this would further reduce the number of adjustable parameters and link the functional even more to the
microscopic approach. Essentially only the spin-orbit and pairing contributions would then need some
phenomenological adjustments. However, the G-matrix also has spin-spin interaction terms and, in principle, the
spin orbit also could be extracted if the  vector part of the one-body density matrix is properly taken into
account \cite{Duguet2}. To what extent such a goal can be achieved remains an open problem. Let us, however,
point to some possible improvements over the past procedures. In first place we want to point out that it is our
believe that neither the DME of Negele- Vautherin, nor the one of Campi-Bouissy \cite{Campi} are the optimal
procedures. We have long standing experience with the semiclassical DME of Wigner and Kirkwood which is based on
a systematic expansion in powers of $\hbar$. Most accurate results are obtained, at least for the scalar part of
the one- body density matrix, with this method \cite{XavierDM}. A second problem stems from the way how the
G-matrix of an infinite matter calculation is used for a finite nucleus. The standard procedure is the LDA, that
is one replaces the $k_F$ dependence by a density dependence via the standard relation $k_F \sim \rho^{1/3}(R)$.
The $k_F$ dependence enters mostly into the Pauli operator $\Theta(p_1^2/2m -k_F)\Theta(p_2^2/2m -k_F)$.
However, dependence also can be in the self consistent single particle energies. We want to point out that just
replacing $k_F$ by $\rho(R)$ is not necessarily the best procedure. Our argument goes as follows. The two
particle propagator entering the G-matrix equation

\begin{equation}
G^0(E;\hat H_1,\hat H_2) = [\Theta(\hat H_1 - \mu)\Theta(\hat H_2 -\mu)] [E - \hat H_1 - \hat H_2 -i\eta]^{-1}
\end{equation}

\noindent has an obvious semiclassical, i.e. $\hbar \rightarrow 0$ limit, namely we just replace the single
particle Hamiltonians $\hat H_i$ by their classical counterparts $H_{cl., i}({\bf R},{\bf p})$ with ${\bf
R},{\bf p}$ the phase space variables of position and momenta. With this approximation the two particles move
according to their respective classical Hamiltonians, each. Supposing for simplicity a local mean field $V({\bf
R})$ (the argument can be generalized to nonlocal mean field potentials), we see that in the semiclassical two
body propagator two local chemical potentials appear: $\mu({\bf R}_1)$ and $\mu({\bf R}_2)$ ($\mu({\bf R}_i) =
\mu - V({\bf R}_i)$) and, therefore, two Fermi momenta and, thus, two densities at positions ${\bf R}_{1,2}$
appear in the semiclassical propagator and correspondingly in the G-matrix. Since the effective range of the
interaction is over 2 fm wide, this non locality in the densities may be of quite some importance. One may call
this the Non Local Density Approximation (NLDA). The standard LDA is recovered in putting ${\bf R}_1 = {\bf R}_2
={\bf R}$. This NLDA effect may be more important, or at least of equal importance, than to keep gradient
expansions of the density itself. It could be interesting to test this. Since finite size effects are surface
effects, one may argue that only relatively low densities will be involved. It has recently been shown, that at
least in neutron matter for low densities a separable force works very well for the G-matrix \cite{Maieron},
reducing very much the numerical effort. So eventually one could combine the NLDA approach with a separable
force to get to a surface term. Of course, one will not imagine that with such a procedure one will hit the good
result right on the spot. But even if the result is only semi quantitatively correct, it still may lead to a
reduction of the open parameters and to more insight into the underlying physics. This may be then a first step
in the direction to get everything from the microscopic input, i.e. from the underlying bare forces, also for
finite nuclei.

\section{Conclusions and prospects}

We have presented a formulation of the energy functional method that enables to keep, to a certain extent, the
connection with the nuclear matter Equation of State as calculated microscopically from many-body theory and
realistic nucleon-nucleon interactions. The functional is constructed following closely the Kohn and Sham
method, where the bulk part of the functional is taken directly from the microscopic calculations and kept fixed
in the fitting procedure of the adjustable parameters of the functional. The latter include the surface part of
the functional and the spin-orbit strength, ending up with a total of four free parameters. The pairing
interaction and the corresponding strength has been taken from the simplest standard choices. Despite the
analysis was not carried out up to the optimal level, leaving space for refinements, the accuracy of the results
can compete with the one obtained with the best purely phenomenological functional like Gogny. These promising
results open the possibility of building an energy density functional closely connected with the bare
nucleon-nucleon forces. Developments in this direction could be obtained along the density matrix expansion
method of ref. \cite{NV72, Duguet2} or its refinements, which could further reduce the number of adjustable
parameters. However, we pointed out the various open problems that must be solved before this project can be
completed. The incorporation of the nuclear EOS in the density functional can be done in different ways, and
this leaves some ambiguity in the direction of a universal nuclear functional. Furthermore the very low density
part of the EOS cannot be taken literarily from microscopic calculations of homogeneous nuclear matter. In fact
cluster formation is expected to dominate this region of the EOS. This appears a serious problem for any density
functional theory which is constructed in such a way to be compatible with the nuclear matter case. Further
studies and analysis are needed to proceed further in this project on the nuclear density functional.

\section{Acknowledgments}
Work supported in part by MICINN (Spain) grants FPA2007-66069 and FPA2008-03865-E/IN2P3, and by the
Consolider-Ingenio 2010 program (Spain) CPAN, CSD2007-00042. X. V. also acknowledges the support from
FIS2008-01661 (Spain and FEDER) and 2009SGR-1289 (Spain). This work was supported by CompStar, a Research
Networking Programme of the European Science Foundation.

{}

\end{document}